\documentclass[final,10pt,twocolumn,oneside,journal]{IEEEtran}

\usepackage{amsmath}
\usepackage{amsthm}
\usepackage{amsbsy}
\usepackage{amssymb}
\usepackage{amscd}
\usepackage{amsfonts}
\usepackage{latexsym}
\usepackage{epsf}
\usepackage{graphicx}
\usepackage{psfrag}
\usepackage[T1]{fontenc}
\usepackage{cite}
\usepackage{booktabs}
\usepackage[mathlines]{lineno}
\usepackage{color}
\usepackage{setspace}

\def\bzero{{\boldsymbol{0}}}
\def\bh{{\boldsymbol{h}}}
\def\bw{{\boldsymbol{w}}}
\def\bH{{\boldsymbol{H}}}
\def\bQ{{\boldsymbol{Q}}}
\def\bPsi{{\boldsymbol{\Psi}}}

\def\mR{{{\mathcal{R}}}}
\def\mS{{{\mathcal{S}}}}
\def\Ab{{{\mathbb{A}}}}
\def\Bb{{{\mathbb{B}}}}
\def\Cb{{{\mathbb{C}}}}
\def\Db{{{\mathbb{D}}}}
\def\Xb{{{\mathbb{X}}}}

\def\ind{{{\text{ind}}}}
\def\com{{{\text{com}}}}
\def\stat{{{\text{stat}}}}
\def\inst{{{\text{inst}}}}

\def\SU{^\text{SU}}

\newcommand{\norm}[1]{{ \left\Vert #1 \right\Vert }}
\newcommand{\snorm}[1]{{ \left\Vert #1 \right\Vert^2 }}

\DeclareMathOperator{\rankof}{rank}
\DeclareMathOperator{\tra}{trace}
\newcommand{\rank}[1]{{ \rankof\!\left\lbrace  #1  \right\rbrace }}
\newcommand{\trace}[1]{{ \tra\!\left\lbrace  #1  \right\rbrace }}

\theoremstyle{plain}
\newtheorem{prop}{Proposition}
\theoremstyle{definition}
\newtheorem{defn}{Definition} 

\IEEEoverridecommandlockouts

\title{Achievable Outage Rate Regions \\ for the MISO Interference 
Channel} 

\author{Johannes~Lindblom, ~\IEEEmembership{Student Member,~IEEE,}
  Eleftherios~Karipidis,~\IEEEmembership{Member,~IEEE,}
  and~Erik~G.~Larsson,~\IEEEmembership{Senior~Member,~IEEE} \vspace*{-5mm}%
  \thanks{
  	The authors are with the Communication Systems Division, Department of 
  	Electrical Engineering (ISY), 
  Link\"oping University, SE-581 83 Link\"oping, Sweden (e-mail:
    \{lindblom,karipidis,erik.larsson\}@isy.liu.se)

    This work has been supported in part by the Swedish Research
    Council (VR), the Swedish Foundation of Strategic Research
    (SSF), and the Excellence Center at Link\"oping-Lund in Information
    Technology (ELLIIT). This work has been performed in the framework
    of the European research project SAPHYRE, which was partly funded by
    the European Union under its FP7 ICT Objective 1.1 - The Network of
    the Future.

    A preliminary version of parts of this work was presented at the Asilomar 
Conference 2009 \cite{Lindblom-2009-Asilomar}.
  }
}

\begin{document}

\maketitle

\begin{abstract}
We consider the slow-fading two-user multiple-input single-output (MISO) 
interference channel. We want to understand which rate points can be 
achieved, allowing a non-zero outage probability. We do so by defining 
four different outage rate regions. The definitions differ on whether the 
rates are declared in outage jointly or individually and whether the 
transmitters have instantaneous or statistical channel state information 
(CSI). The focus is on the instantaneous CSI case with individual 
outage, where we propose a stochastic mapping from the rate point and the 
channel realization to the beamforming vectors. A major contribution is 
that we prove that the stochastic component of this mapping is independent of 
the actual channel realization.
\end{abstract}


\section{Introduction}\label{sec:intro}
\IEEEPARstart{W}{e} study the two-user multiple-input single-output (MISO) 
interference channel 
(IC), consisting of two transmitter (TX) - receiver (RX) pairs (or links). The 
transmissions are concurrent and cochannel; hence, they interfere with each 
other. The TXs employ multiple antennas and the RXs a single antenna. We 
assume that the channels are flat and slow fading and we say that a link is in 
\emph{outage} if the IC experiences fading states that cannot support a 
desired data rate. The fundamental question raised is how to define the 
\emph{outage rate region}. That is, which rate points can be achieved with a 
certain probability? For multi-user systems, such as the IC, broadcast channel 
(BC), and multiple-access channel (MAC), one can consider \emph{common} or 
\emph{individual} outage. We declare a common outage if the rate of at least 
one link cannot be supported (see, e.g., \cite{Li-2001-TIT} for the BC). We 
declare an individual outage if a specific link is unable to communicate at 
the desired rate.

So far, studies of outage rate regions have been restricted to the 
single-antenna BC and MAC for which the outage capacity regions for 
instantaneous channel side information (CSI) were given in \cite{Li-2001-TIT} 
and \cite{Li-2005-TIT}, respectively. For statistical CSI, the MAC and BC were 
studied in \cite{Narasimhan-2007-ISIT} and \cite{Zhang-2009-TIT}, 
respectively. The instantaneous rate region for the MISO IC 
is well-understood (see, e.g., \cite{Bjornson-2010-TSP} and 
\cite{Mochaourab-2012-JSTSP}). 
In \cite{Lindblom-2009-Asilomar}, we defined the regions for individual and 
common outage for statistical CSI and common outage for instantaneous CSI. 
For the Gaussian IC in the high signal-to-noise ratio (SNR) regime, recent 
research activities have explored the diversity-multiplexing trade-off (DMT) 
(see e.g., \cite{Ebrahimzad-2012-TIT} for characterization of the two-user IC).
In \cite{Li-2013-TSP}, our results in \cite{Lindblom-2009-Asilomar} were used 
to approximately perform weighted sum-rate maximization under outage 
constraints for the MISO IC with statistical CSI. Also for statistical CSI, 
outage probabilities in the multiple-input multiple-output (MIMO) IC were 
given in closed form and an outage-based robust beamforming algorithm was 
proposed in \cite{Park-2012-TWC}.

In this paper, we propose and analyze achievable outage rate regions for the 
MISO IC. The results generalize those of \cite{Li-2001-TIT, Li-2005-TIT, 
Narasimhan-2007-ISIT, Zhang-2009-TIT} both in the sense that the BC and MAC 
are special cases of the IC, and in that we treat the multiple antenna case. 
Since we allow a non-zero outage probability, our results 
extend those of \cite{Bjornson-2010-TSP} and \cite{Mochaourab-2012-JSTSP} 
where outage was not allowed. In contrast to the 
DMT analysis, e.g., \cite{Ebrahimzad-2012-TIT}, our results are valid for any 
SNR regime. For completeness, we consider common and individual outage for 
both instantaneous and statistical CSI, but we focus on the individual 
outage rate region for instantaneous CSI, which we did not treat in 
\cite{Lindblom-2009-Asilomar}. A challenge is how to handle the 
scenario where either of the rates can be achieved, but not simultaneously. We 
solve this by proposing a stochastic mapping of the beamforming vectors that 
depends on the rates and the channels. We prove that the randomness of the 
mapping is independent of the channel realization. Compared to 
\cite{Lindblom-2009-Asilomar}, the statistical CSI definitions extend the 
single-stream transmission scheme to multi-stream. The definitions are valid 
for arbitrary assumptions on the channel distribution. 

\section{System Model}
\label{sec:model}
We assume that the RXs treat the interference as additive Gaussian noise. 
Also, RX$_i$ experiences additive Gaussian thermal noise with variance 
$\sigma_i^2$.
TX$_i$ employs $n$ antennas and uses a Gaussian vector codebook with
covariance $\bPsi_i$. By $\bh_{ij} \in \mathbb{C}^n$, we denote the 
slow-fading conjugated channel vector between TX$_i$ and RX$_j$ and we assume 
that the 
channels $\{\bh_{ij}\}_{i,j=1}^2$ are statistically independent. We let $\bh$ 
denote a specific realization of the channels, i.e., 
$\bh = [\bh_{11}^T, \bh_{12}^T, \bh_{21}^T, \bh_{22}^T]^T$. By $\bH$ we denote 
a random channel with pdf $f_\bH(\bh).$ The achievable rate, in
bits per channel use, of link $i$ is 
\begin{equation*}
R_i(\bh,\bPsi_i,\bPsi_j) = \log_2 \left(1 +
\frac{\bh^H_{ii} \bPsi_i \bh_{ii}} {\bh^H_{ji} \bPsi_j \bh_{ji} +
\sigma_i^2} \right).
\end{equation*}

We bound the transmit power to $\trace{\bPsi_i} \leq 1.$ For statistical
CSI, multi-stream transmission, i.e., $\rank{\bPsi_i} \leq n$ is optimal in
general. However, for instantaneous CSI, single-stream transmission, i.e.,
$\rank{\bPsi_i} = 1$, is optimal \cite{Shang-2011-TIT}. For the latter case,
we set $\bPsi_i = \bw_i \bw_i^H$, where $\bw_i$ is the beamforming vector with
$\snorm{\bw_i} \leq 1.$ For instantaneous CSI, the corresponding rate is denoted
$R_i(\bh,\bw_1,\bw_2)$.

\section{Outage Rate Region for Instantaneous CSI}
\label{sec:inst}

We assume that the TXs have instantaneous CSI and therefore can adapt the
beamforming vectors to the current fading state. 
The definition of the common outage rate region given in Sec. 
\ref{sec:inst_com}, was first proposed in \cite{Lindblom-2009-Asilomar}. Here, 
we give it for completeness. The definition of the individual outage rate 
region is novel and is given in Sec. \ref{sec:inst_ind}.

For a given channel realization $\bh$, the achievable instantaneous rate 
region is the set of rate points that can be achieved by using any pair of 
feasible beamforming vectors, i.e.,
\begin{equation*} 
\mR(\bh) \triangleq \!\!\! \bigcup_{\snorm{\bw_i} \leq 1, i=1,2} \!\!\!
  \left( R_1(\bh, \bw_1, \bw_2), R_2(\bh, \bw_1,\bw_2)
\right).
\end{equation*}
By $R\SU_i(\bh) \triangleq \log_2 ( 1+\snorm{\bh_{ii}} / \sigma_i^2 ),$
we denote the single-user (SU) rate for link $i$, i.e., the maximum 
rate achieved in absence of interference when TX$_i$ uses its matched-filter 
(MF) beamforming vector $\bw_i^\text{MF} \triangleq
\bh_{ii}/\norm{\bh_{ii}}.$

\subsection{Common Outage Rate Region for Instantaneous CSI}\label{sec:inst_com}

We denote by $\mR_\inst^\com (\epsilon)$ the sought common outage rate region
for instantaneous CSI. If a rate point $(r_1,r_2)\notin \mR(\bh)$, i.e., it is
not achievable, we say that the IC is in outage.
\begin{defn} \label{def:CSI_com}
  Let $\epsilon>0$ denote the common outage probability specification. Then,
$(r_1, r_2)\in \mR^\com_\inst(\epsilon)$ if there exists a
deterministic mapping $(\bw_1(\bh,r_1,r_2), \bw_2(\bh,r_1,r_2))$ 
with $\snorm{\bw_i} \leq 1,$ $i=1,2,$ such that 
$\Pr\{r_1 < R_1(\bH,\bw_1, \bw_2), r_2 < R_2(\bH,\bw_1, \bw_2) \} \geq 
1-\epsilon.$
\end{defn}

To determine if $(r_1,r_2)$ is achievable for a channel realization $\bh$,
we can solve a scalar, quasi-concave rate maximization problem that takes 
$r_1$ as input and returns $r_2^\star$~\cite{Lindblom-2012-TSP}. If $r_2^\star 
\geq r_2$, then $(r_1,r_2)$ is achievable and the solution of the rate 
maximization problem gives us the enabling beamforming 
vectors.\footnote{By the symmetry of the problem, we can 
equivalently choose $r_2$ as input and $r_1^\star$ as output of the 
optimization. Then $(r_1,r_2)$ is feasible if $r_1 \leq r_1^\star$.} Note that 
the beamforming vectors, which depend on both the 
channel realization and the rate point, are not unique. Equivalently to Def. 
\ref{def:CSI_com}, we say that $(r_1, r_2)\in \mR^\com_\inst(\epsilon)$ if 
$\Pr\{ (r_1,r_2) \in \mR(\bH) \} \geq 1-\epsilon.$
Since $\Pr\{ (r_1,r_2) \in \mR(\bH) \}$ decreases with respect to one rate 
when the other is fixed, a point on the outer boundary of $\mR_\inst^\com 
(\epsilon)$ has an outage probability equal to $\epsilon$. 

\subsection{Individual Outage Rate Region for Instantaneous 
CSI}\label{sec:inst_ind}
We denote by $\mR_\inst^\ind(\epsilon_1, \epsilon_2)$ the sought individual
outage rate region for instantaneous CSI. In contrast to the common outage
rate region, we assume that when the rate of one link cannot be achieved, the
corresponding TX is switched off. In such occasion, the other link does not 
experience interference, hence it has increased chances of achieving its 
desired rate. 
\begin{defn} \label{def:IndOutageInstCSI}
Let $\epsilon_1, \epsilon_2>0$ denote the individual outage
probability
specifications. Then, $(r_1, r_2)\in \mR^\ind_\inst(\epsilon_1,
\epsilon_2)$ if there exists a stochastic mapping $(\bw_1(\bh,r_1,r_2),
\bw_2(\bh,r_1,r_2))$ with $\snorm{\bw_i} \leq 1,$ $i=1,2,$ such 
that $\Pr \{r_1 < R_1(\bH,\bw_1,\bw_2)\} \geq
1-\epsilon_1$ and $\Pr \{r_2 < R_2(\bH,\bw_1,\bw_2)\} \geq 1-\epsilon_2.$ 
\end{defn}
In the following, we motivate Def. \ref{def:IndOutageInstCSI} by proposing a
stochastic mapping $(\bw_1(\bh,r_1,r_2), \bw_2(\bh,r_1,r_2))$ and conditions for
having $(r_1, r_2)\in \mR^\ind_\inst(\epsilon_1, \epsilon_2)$. 
First, we focus on a given rate point $(r_1,r_2)$ and a realization of the
channels and determine whether the rates $r_1$ and $r_2$ are achievable or not
and outline the stochastic mapping. Either none of the rates is achievable, or
both of
them are achievable, or only one of them is achievable. We formalize this by
serially performing the following checks:

1) Is $r_1>R\SU_1(\bh)$ and $r_2>R\SU_2(\bh)$? If yes, we have case 
$\Ab$: none of
$r_1$ and $r_2$ is achievable, we set $\bw_1 = \bw_2 = \bzero,$ and both links
are in outage. 

2) Is $(r_1,r_2) \in \mR(\bh)$? If yes, we have case $\Bb$. We find
$(\bw_1,\bw_2)$ by solving the rate maximization problem in
\cite{Lindblom-2012-TSP}. Note that this is the only case where 
the MISO IC is not in outage under the common outage specification.

3) Is $r_1>R\SU_1(\bh)$ or $r_2>R\SU_2(\bh)$? If $r_2>R\SU_2(\bh)$, we have 
case $\Cb_1$: $r_1$ is achievable with $\bw_1 = \bw_1^\text{MF}$ while $r_2$ 
is in outage and we set $\bw_2 = \bzero$. If $r_1>R\SU_1(\bh)$, we have case 
$\Cb_2$: $r_2$ is achievable with $\bw_2 = \bw_2^\text{MF}$ while $r_1$ is in 
outage and we set $\bw_1 = \bzero$. If neither $r_1>R\SU_1(\bh)$ nor 
$r_2>R\SU_2(\bh)$, we have case $\Db$: there is an ambiguity; both rates can 
be achieved, but not simultaneously. Here, we decide that one of the links is 
active while the other is in outage. For example, we could always decide in 
favor of link 1. But, as we illustrate in Sec. \ref{sec:numerical} and 
Fig.~\ref{fig:regions}, 
this is a suboptimal strategy. In order to have as large region as possible, 
we make this binary decision at random, e.g., by flipping a biased coin. 
We let $I\in\{1,2\}$ be the outcome of the coin flip and assume that 
the coin has bias $\Pr \{I=1 | \bH = \bh; r_1,r_2\} = 1-\Pr \{I=2 | \bH = \bh; 
r_1,r_2\}$. That is, the coin bias depends on both the channel realization 
and the specific rate point.

4) Is $I=1$? We have case $\Db_1$: we achieve $r_1$ while $r_2$ is in outage 
by using the same beamforming vectors as in $\Cb_1$. Is $I=2$? We have case 
$\Db_2$: we achieve $r_2$ while $r_1$ is in outage by using the same 
beamforming vectors as in $\Cb_2$. 

Second, we consider the entire set of channel realizations and a specific rate
point $(r_1,r_2)$. We define $\mS_\Xb(r_1,r_2)$ to be the set of channel 
realizations for which $(r_1,r_2)$ falls under case $\Xb \in 
\{\Ab,\Bb,\Cb_1,\Cb_2,\Db\}$. Therefore, we have
\begin{align}
\mS_\Ab(r_1,r_2) \triangleq & \{\bh: r_1 > R\SU_1(\bh), r_2 > R\SU_2(\bh)\},
\label{eq:SA} \\
\mS_\Bb(r_1,r_2) \triangleq & \{\bh: (r_1,r_2) \in \mR(\bh)\}, \label{eq:SB} \\
\mS_{\Cb_1}(r_1,r_2) \triangleq & \{\bh: r_1 \leq R\SU_1(\bh), r_2 > 
R\SU_2(\bh)\}, \label{eq:SC1} \\
\mS_{\Cb_2}(r_1,r_2) \triangleq & \{\bh: r_1 > R\SU_1(\bh), r_2 \leq 
R\SU_2(\bh)\}, \label{eq:SC2} \\
\mS_\Db(r_1,r_2) \triangleq & \{\bh: r_1 \leq R\SU_1(\bh), r_2 \leq 
R\SU_2(\bh), \nonumber \\
&(r_1,r_2) \notin \mR(\bh) \}. \label{eq:SD} 
\end{align} 
Since the events $\Ab,$ $\Bb,$ $\Cb_1,$ $\Cb_2,$ and $\Db$ are mutually 
exclusive, we can conclude that the sets defined in 
\eqref{eq:SA}--\eqref{eq:SD} are disjoint and span the set of all channel 
realizations.

The probability that a channel realization belongs to each of the sets
defined in \eqref{eq:SA}--\eqref{eq:SD} is $ P_\Xb(r_1,r_2) \triangleq \Pr \{
\bH \in \mS_\Xb(r_1,r_2) \} $ for $\Xb \in \{\Ab,\Bb,\Cb_1,\Cb_2,\Db\}.$
Especially, we have 
\begin{align}
  P_\Ab(r_1,r_2) = & \Pr\{r_1 > R\SU_1(\bH)\} \Pr \{r_2 > 
R\SU_2(\bH)\},
  \label{eq:PrA}\\
  P_\Bb(r_1,r_2) = & \Pr \{ (r_1,r_2) \in \mR(\bH) \}, \label{eq:PrB} \\
  P_{\Cb_1}(r_1,r_2) = & \Pr\{r_1 \leq R\SU_1(\bH)\} \Pr \{r_2 > 
R\SU_2(\bH)\},
  \label{eq:PrC1}\\
  P_{\Cb_2}(r_1,r_2) = & \Pr\{r_1 > R\SU_1(\bH) \} \Pr\{r_2 \leq 
R\SU_2(\bH) \},
  \label{eq:PrC2} \\
  P_{\Db}(r_1,r_2) = & \Pr\{r_1 \leq R\SU_1(\bH) \} \Pr\{r_2 \leq 
R\SU_2(\bH) \}
\nonumber \\ 
  &-  \Pr \{ (r_1,r_2) \in \mR(\bH) \},  \label{eq:PrD}
\end{align}
where \eqref{eq:PrA}, \eqref{eq:PrC1}, and \eqref{eq:PrC2} follow by the 
independence of the random variables $R\SU_1(\bH)$ and $R\SU_2(\bH)$. For 
\eqref{eq:PrD} we use that
\begin{align*}
	&\Pr \{r_1 \leq R\SU_1(\bH), r_2 \leq R\SU_2(\bH) \} \nonumber \\
	& = \Pr \{r_1 \leq R\SU_1(\bH), r_2 \leq R\SU_2(\bH) | (r_1,r_2) 
	\in \mR(\bH)\} \nonumber \\
	& ~~~~~~\times \Pr \{(r_1,r_2) \in \mR(\bH) \} \nonumber \\
	& ~~~ + \Pr \{r_1 \leq R\SU_1(\bH), r_2 \leq R\SU_2(\bH), (r_1,r_2) 
	\notin \mR(\bH)\} \nonumber \\
	& = \Pr \{(r_1,r_2) \in \mR(\bH)\} + P_{\Db}(r_1,r_2),
\end{align*}
since $\Pr \{r_1 \leq R\SU_1(\bH), r_2 \leq R\SU_2(\bH) | (r_1,r_2) \in 
\mR(\bH)\} = 1$. 
It is straightforward to verify that the probabilities in
\eqref{eq:PrA}--\eqref{eq:PrD} sum up to one. 
For case $\Db_i$, we introduce the joint mixed distribution $f_{\bH,I} 
(\bh,i;r_1,r_2) = \Pr\{I=i|\bH=\bh; r_1,r_2\} f_\bH(\bh)$. The interpretation 
is that the coin bias depends on both the realization of the channels and the 
rate point. We have 
\begin{align}
&P_{\Db_i}(r_1,r_2) \triangleq \Pr\{ \bh \in \mS_\Db(r_1,r_2), I = i;r_1,r_2 \}
\nonumber \\ 
&= \int_{\mS_\Db(r_1,r_2)} \Pr\{I=i | \bH = \bh; r_1,r_2\} f_\bH(\bh) d\bh.
\label{eq:ProbDi}
\end{align}
Based on the discussion above, we write the outage constraint for link $i$ in 
Def.~\ref{def:IndOutageInstCSI} as
\begin{align}
&\Pr\{R_i(\bH, \bw_1, \bw_2) > r_i\} \nonumber \\ &=
P_\Bb(r_1,r_2) + P_{\Cb_i}(r_1,r_2) + P_{\Db_i}(r_1,r_2) \geq 1-\epsilon_i.
\label{eq:usagei}
\end{align}

Third, we focus on the coin bias $\Pr \{I=i | \bH = \bh; r_1,r_2 \}$.
\begin{prop} \label{prop:bias}
The coin bias can be chosen independently of the realization of the channels,
i.e., $\Pr\{I=1 | \bH = \bh;r_1,r_2\} = p(r_1,r_2)$ and $\Pr\{I=2 | \bH = \bh;
r_1,r_2\} = 1-p(r_1,r_2)$.
\end{prop}
\begin{IEEEproof} 
	According to Def.~\ref{def:IndOutageInstCSI}, $(r_1, r_2)\in 
	\mR^\ind_\inst(\epsilon_1, \epsilon_2)$ if \eqref{eq:usagei} is satisfied 
	for $i=1,2.$ By inserting \eqref{eq:ProbDi} into~\eqref{eq:usagei}, we get 
	the equivalent conditions
\begin{align}
 & 1-\epsilon_1 - P_\Bb(r_1,r_2) - P_{\Cb_1}(r_1,r_2) \leq P_{\Db_1}(r_1,r_2) 
 \nonumber \\
 & = \int_{\mS_\Db(r_1,r_2)} \Pr\{I=1 | \bH = \bh;r_1,r_2\} f_\bH(\bh) d\bh 
\nonumber \\
 &\leq P_\Bb(r_1,r_2) + P_{\Cb_2}(r_1,r_2) + P_\Db(r_1,r_2) -1+\epsilon_2
\label{eq:usageConstraint},
\end{align}
where the last inequality follows since $P_{\Db_1}(r_1,r_2) =  
P_{\Db}(r_1,r_2) 
- P_{\Db_2}(r_1,r_2).$ By selecting $\Pr\{I=1 | \bH = \bh;r_1,r_2\}$ 
appropriately, we can force the integral in \eqref{eq:usageConstraint} to 
assume any value in $[0,P_\Db(r_1,r_2)].$ If we restrict $\Pr\{I=1 | \bH 
= \bh;r_1,r_2\} = p(r_1,r_2)$ for some function of $(r_1,r_2)$ that does not 
depend on $\bH,$ then likewise, we can force the integral to assume any value 
in $[0,P_\Db(r_1,r_2)]$ as well.
\end{IEEEproof}

By using the result of Prop.~\ref{prop:bias}, we can write 
\eqref{eq:usageConstraint} as
\begin{align}
	& 1-\epsilon_1 - P_\Bb(r_1,r_2) - P_{\Cb_1}(r_1,r_2)
	\leq P_\Db(r_1,r_2) p(r_1,r_2) \nonumber \\
	&\leq P_\Bb(r_1,r_2) + P_{\Cb_2}(r_1,r_2) + P_\Db(r_1,r_2) -
		1+\epsilon_2 \label{eq:pConstraint}
\end{align}
If we can find a bias $p(r_1,r_2)\in[0,1]$ which satisfies 
\eqref{eq:pConstraint}, then we have $(r_1,r_2) \in \mR^\ind_\inst 
(\epsilon_1, \epsilon_2)$. 
In order to have $0 \leq p(r_1,r_2) \leq 1,$ the following 
conditions must be satisfied: a) The lower bound in \eqref{eq:pConstraint} is 
less than  $P_\Db(r_1,r_2)$. b) The upper bound is non-negative. c) The lower 
bound is smaller 
than the upper bound. Hence, using the fact that the probabilities 
\eqref{eq:PrA}--\eqref{eq:PrB} sum up to one, we have the conditions
\begin{align}
	\epsilon_1 & \geq P_\Ab(r_1,r_2) + P_{\Cb_2}(r_1,r_2), \label{eq:OutReg1}
\\ 
	\epsilon_2 & \geq P_\Ab(r_1,r_2) + P_{\Cb_1}(r_1,r_2), \label{eq:OutReg2}
\\
	\epsilon_1 + \epsilon_2 & \geq 1+P_\Ab(r_1,r_2) - P_\Bb(r_1,r_2).
\label{eq:OutRegSum}
\end{align}
If all conditions in \eqref{eq:OutReg1}--\eqref{eq:OutRegSum} are satisfied, we
choose $p(r_1,r_2)$ according to \eqref{eq:pConstraint} and the rate point
$(r_1,r_2)$ lies in the individual outage rate region. Otherwise, $(r_1,r_2)$
does not belong to the outage rate region. To give some
interpretation, we insert \eqref{eq:PrA}--\eqref{eq:PrC2} into
\eqref{eq:OutReg1}--\eqref{eq:OutRegSum}, and get
\begin{align}
\epsilon_1 \geq & \Pr\{r_1 > R\SU_1(\bH)\} \label{eq:OutReg1_elab}, \\
\epsilon_2 \geq & \Pr\{r_2 > R\SU_2(\bH)\} \label{eq:OutReg2_elab}, \\
\epsilon_1 + \epsilon_2 \geq & \Pr \{ r_1 > R\SU_1(\bH)\} \Pr\{r_2 > 
R\SU_2(\bH) \}
\nonumber \\ &+ \Pr \{ (r_1,r_2) \notin \mR(\bH) \}.
\label{eq:OutRegSum_elab}
\end{align}
It is apparent that $\Pr\{r_1 > R\SU_1(\bH)\}$ and $\Pr\{r_2 > R\SU_2(\bH)\}$ 
are decreasing with $r_1$ and $r_2,$ respectively. Also, $\Pr \{ (r_1, r_2) 
\notin \mR(\bH) \}$ increases when one of the rates increases but the other 
is fixed. Therefore, we conclude that points on the outer boundary of the 
outage rate region must satisfy at least one of the inequalities 
\eqref{eq:OutReg1_elab}--\eqref{eq:OutRegSum_elab} with equality. Another 
observation is that \eqref{eq:OutReg1_elab} and 
\eqref{eq:OutReg2_elab} are the trivial outage constraints for the SU points, 
i.e., SU MISO channel, whereas \eqref{eq:OutRegSum_elab} gives 
the shrinkage of the outage rate region due to interference. Note that, 
equivalently to Def.~\ref{def:IndOutageInstCSI}, we can define $\mR^\ind_\inst 
(\epsilon_1, \epsilon_2)$ as the set of rate points which satisfy 
\eqref{eq:OutReg1_elab}--\eqref{eq:OutRegSum_elab}.

\section{Outage Rate Regions for Statistical CSI}\label{sec:stat}

We assume that the TXs only have knowledge of the channels' statistical 
distribution. That is, the TXs have statistical CSI and can only adapt their 
transmit covariance matrices to the channel statistics. Therefore, the 
TXs design the transmit covariance matrices once and use them for all fading 
states. The definitions are given for completeness; for details we refer to 
\cite{Lindblom-2009-Asilomar}. We give definitions for the common and 
individual outage rate regions in Secs. \ref{sec:stat_com} and 
\ref{sec:stat_ind}, respectively. 

\subsection{Common Outage Rate Region for Statistical CSI}\label{sec:stat_com}

We denote by $\mR_\stat^\com (\epsilon)$ the sought common outage rate region
for statistical CSI and define it as follows.
\begin{defn}
\label{def:CDI_com}
Let $\epsilon > 0$ denote the {\it common} outage probability specification.
Then, $(r_1, r_2)\in \mR_\stat^\com (\epsilon)$ if there exists
a deterministic mapping $(\bPsi_1(f_\bH(\bh),r_1, r_2),\bPsi_2(f_\bH(\bh),r_1, 
r_2))$ with $\trace{\bPsi_i} \leq 1$ for $i=1,2$ such that
$
  \Pr \left\{ R_1(\bH,\bPsi_1, \bPsi_2) > r_1, R_2(\bH,\bPsi_1, \bPsi_2) > r_2
 \right\} \geq 1-\epsilon.
$ 
\end{defn}
Note that the transmit covariance matrices $\bPsi_1$ and $\bPsi_2$ depend on 
both the channel statistics and the actual rate point.

\subsection{Individual Outage Rate Region for Statistical
CSI}\label{sec:stat_ind}
We denote by $\mR_\stat^\ind(\epsilon_1,\epsilon_2)$ the sought individual 
outage rate region for statistical CSI. We allow one link to be in outage 
while the other is not. Since the TXs do not know whether the transmission 
is in outage or not, a TX continues transmitting even when the 
link is in outage.

\vspace*{-2mm}
\begin{defn}
\label{def:CDI_ind}
Let $\epsilon_1, \epsilon_2 > 0$ denote the {\it individual} outage
probability specifications. Then, $(r_1,r_2) \in \mR_\stat^\ind (\epsilon_1,
\epsilon_2)$ if there exists a deterministic mapping 
$(\bPsi_1(f_\bH(\bh),r_1,r_2),\bPsi_2(f_\bH(\bh),r_1,r_2)$ with  
$\trace{\bPsi_i} \leq 1$ for $i=1,2$ such that
$
\Pr\{R_i(\bH, \bPsi_1, \bPsi_2) \geq r_i \} \geq 1-\epsilon_i.
$ 
\end{defn}

\begin{figure}[t]
	\vspace*{-4mm}
  \begin{center}
    \psfrag{Common outage, statistical CSI}[l][l][1]{Common outage, 
statistical CSI}
    \psfrag{Individual outage, statistical CSI}[l][l][1]{Individual outage,
statistical CSI}
    \psfrag{Common outage, instantaneous CSI}[l][l][1]{Common outage, 
instantaneous CSI}
    \psfrag{Individual outage, instantaneous CSI}[l][l][1]{Individual outage,
  instantaneous CSI}
  \psfrag{Choose link 1}[l][l][1]{$\mR_\inst^\ind(\epsilon_1,\epsilon_2)$ for 
$p(r_1,r_2)=1$}
  \psfrag{Choose link 2}[l][l][1]{$\mR_\inst^\ind(\epsilon_1,\epsilon_2)$ for 
$p(r_1,r_2)=0$}
    \psfrag{R1 [bpcu]}[c][c][1.2]{$R_1$ [bits/channel use]}
    \psfrag{R2 [bpcu]}[c][c][1.2]{$R_2$ [bits/channel use]}
    \psfrag{0}[c][c][1]{0}
    \psfrag{0.2}[c][c][1]{0.2}
    \psfrag{0.4}[c][c][1]{0.4}
    \psfrag{0.6}[c][c][1]{0.6}
    \psfrag{0.8}[c][c][1]{0.8}
    \psfrag{1}[c][c][1]{1}
    \psfrag{1.2}[c][c][1]{1.2}
    \psfrag{1.4}[c][c][1]{1.4}
    \psfrag{2}[c][c][1]{2}
    \psfrag{3}[c][c][1]{3}
    \psfrag{4}[c][c][1]{4}
    \begin{psfrags}
      \resizebox{0.95\linewidth}{!}{\epsfbox{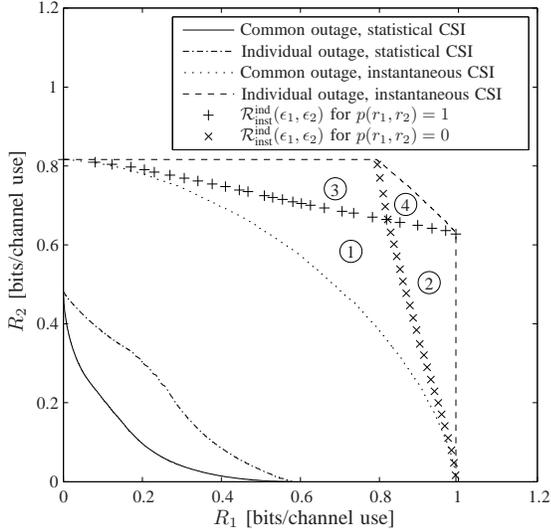}}
    \end{psfrags}
  \end{center}
  \vspace*{-7mm}
  \caption{Outer boundaries of the outage rate regions for the MISO IC. Also, 
we illustrate the effect of choosing optimal bias for the
  	individual outage rate region for instantaneous CSI.}
  \vspace*{-4mm}
  \label{fig:regions}
\end{figure}

\vspace*{-2mm}
\section{Numerical Example}\label{sec:numerical}
  
We illustrate the outage rate regions given in 
Defs.~\ref{def:CSI_com}--\ref{def:CDI_ind}. The TXs employ $n=2$ antennas 
each and we model $\bh_{ij}\in\Cb^n$ as a
zero-mean complex-symmetric Gaussian vector with covariance $\bQ_{ij}$. 
We assume that $\sigma_1^2 = \sigma_2^2 = 0.5$ and $\epsilon = \epsilon_1 =
\epsilon_2 = 0.1$. For a given set of channel covariance matrices 
$\{\bQ_{ij}\}$, we depict the regions in Fig.~\ref{fig:regions}.

We use exhaustive-search methods to generate the regions. For instantaneous 
CSI, we make a grid of rate points. Then, for each rate point, we estimate the 
probabilities \eqref{eq:OutReg1_elab}--\eqref{eq:OutRegSum_elab} by running 
Monte-Carlo simulations. For determining if $(r_1,r_2) \in \mR(\bh),$ we use 
the fast method given in \cite{Lindblom-2012-TSP}. For statistical CSI we draw 
beamforming vectors randomly. Using results from \cite{Bletsas-2007-TWC}, we 
compute the probabilities in Defs.~\ref{def:CDI_com} and~\ref{def:CDI_ind} in 
closed form. For each pair of beamforming vectors, we determine the rate 
points that meet the outage specifications. We find the outer boundary via a 
brute-force comparison among all computed rate points.

We observe that the individual outage regions are larger than the 
corresponding common outage regions and the instantaneous CSI regions are 
larger than the corresponding statistical CSI regions. These results are 
expected since common outage is more restrictive than individual outage and 
instantaneous CSI is always better than statistical CSI. These results are 
true in general, but we omit the proof due to space limitations.

We also illustrate the effect of choosing the bias according to 
\eqref{eq:pConstraint}. Area 1 is the gain, compared to the common outage 
case, from including the obvious cases $\Cb_1$ and $\Cb_2$. Area 2 (or 3) is 
the gain from solving the conflict by always choosing in favor of link 1 (or 
2), i.e., by switching off deterministically. Area 4 is the gain from randomly 
switching off the transmissions using the bias according 
to~\eqref{eq:pConstraint}. 

\section{Conclusions}\label{sec:conc}

We defined four outage rate regions for the MISO IC. The definitions 
correspond to different scenarios of channel knowledge and outage 
specification. We observe that neither the definitions depend on the channels' 
distributions nor they are restricted for Gaussian coding. On the other hand, 
for Gaussian coding and channels, we have efficient methods for illustrating 
the regions. Whereas the definitions for statistical CSI
assume that interference is treated as noise, the definitions for
instantaneous CSI are valid for any achievable rate region and could
potentially be extended to the MIMO IC.

\bibliographystyle{IEEEbib}

\end{document}